\documentclass[manuscript]{biometrika}
\usepackage{dsfont}
\usepackage{amsmath}
\usepackage{amsfonts}
\usepackage{times}
\usepackage{bm}
\usepackage{natbib}
\usepackage{graphicx}
\usepackage{subfigure}
\usepackage{cases}
\usepackage{multirow}
\usepackage[plain,noend]{algorithm2e}

\makeatletter
\renewcommand{\algocf@captiontext}[2]{#1\algocf@typo. \AlCapFnt{}#2} 
\def\@algocf@capt@plain{top}
\renewcommand{\algocf@makecaption}[2]{%
  \addtolength{\hsize}{\algomargin}%
  \sbox\@tempboxa{\algocf@captiontext{#1}{#2}}%
  \ifdim\wd\@tempboxa >\hsize
    \hskip .5\algomargin%
    \parbox[t]{\hsize}{\algocf@captiontext{#1}{#2}}
  \else%
    \global\@minipagefalse%
    \hbox to\hsize{\box\@tempboxa}
  \fi%
  \addtolength{\hsize}{-\algomargin}%
}
\makeatother

\newcommand{\nop}[1]{}

\begin{document}

\jname{Biometrika}
\jyear{*}
\jvol{*}
\jnum{*}
\accessdate{*}
\copyrightinfo{\Copyright\ 2012 Biometrika Trust\goodbreak {\em Printed in Great Britain}}

\received{April 2012}
\revised{September 2012}

\markboth{HUILI YUAN, RUIBIN XI AND MINGHUA DENG}{Differential Network Analysis via Lasso Penalized D-Trace Loss}

\title{Differential Network Analysis via Lasso Penalized D-Trace Loss}

\author{Huili Yuan}
\affil{School of Mathematical Sciences, Peking University, Beijing, P. R. China \email{hlyuan@pku.edu.cn} }

\author{\and Ruibin Xi}
\affil{School of Mathematical Sciences and Center for Statistical Science, Peking University, Beijing, P. R. China \email{ruibinxi@math.pku.edu.cn}}

\author{\and Chong Chen}
\affil{School of Mathematical Sciences and Center for Statistical Science, Peking University, Beijing, P. R. China \email{cheung1990@126.com}}

\author{\and Minghua Deng}
\affil{School of Mathematical Sciences and Center for Statistical Science, Peking University, Beijing, P. R. China \email{dengmh@math.pku.edu.cn}}

\maketitle

\begin{abstract}
Biological networks often change under different environmental and genetic conditions. Understanding how these networks change becomes an important problem in biological studies. In this paper, we model the network change as the difference of two precision matrices and propose a novel loss function called the D-trace loss. Compared to other methods, this D-trace loss function allows us to directly estimate the precision matrix difference without attempting to estimate precision matrices. Under a new irrepresentability condition, we show that the D-trace loss function with the lasso penalty can give consistent estimates in high-dimensional setting if the difference network is sparse. A very efficient algorithm is developed based on the alternating direction method to minimize the lasso penalized D-trace loss function. Simulation studies and a real data analysis about colorectal cancer show that the proposed method outperforms other available methods.
\end{abstract}


\section{Introduction}

\label{sec1}
 Network approaches have been widely used to study interactions of molecular entities such as mRNAs, proteins and microRNAs \citep{basso2005reverse,bonneau2007predictive,pereira2004detection,zhang2012inferring,leiserson2014pan}. It is known that these interactions can change under various environmental and genetic conditions \citep{songyang1995catalytic,bandyopadhyay2010rewiring}, but most netwwork methods were developed for single static condition \citep{ideker2012differential}. Gene regulatory networks are often modelled with Gaussian graphical model \citep{markowetz2007inferring}, where the gene expressions are assumed to be jointly Gaussian and two genes have interaction if and only if the corresponding entry of the precision matrix is nonzero. In this paper, we also model the network as the precision matrix, but we are interested in the difference between two precision matrices. More specifically, suppose that we have independent observations of $p$ genes from two groups of subjects: $X_i=(X_{i1},\ldots,X_{ip})^T$ for $ i=1,\ldots, n_X$ from group 1 and $Y_i=(Y_{i1},\ldots,Y_{ip})^T$ for $ i=1,\ldots, n_Y$ from group 2. The two groups can correspond to two different environmental conditions or two different genetic conditions. Assume that the covariance matrices for group 1 and 2 are $\Sigma_X^*=(\Sigma_{X,ij}^*)$, $\Sigma_Y^*=(\Sigma_{Y,ij}^*)$, respectively. The differential network is defined as the difference between two precision matrices, denoted by $\Delta^*=(\Sigma_Y^*)^{-1}-(\Sigma_X^*)^{-1}$.

There have been active researches on precision matrix estimation in high dimensional setting in recent years \citep{yuan2007model,meinshausen2006high,cai2011constrained,zhang2014sparse}.  A key assumption of these methods is that the precision matrix is sparse and hence one can recover the precision matrix in high dimensional setting. These sparse precision matrix estimation methods are generally not directly applicable to differential network analysis. Firstly, the precision matrices may not be sparse, simply taking difference between two estimated precision matrices would generate many false positives and negatives. Even if the precision matrices are sparse, these sparse precision matrix estimation methods are most powerful in detecting strong interactions in a single static condition, and they will have limited power in detecting interactions that are not strong in a static condition but have large changes in different conditions.

Several methods have been proposed to estimate the differential network. One class of methods \citep{guo2011joint,chiquet2011inferring,danaher2014joint} jointly estimate precision matrices and their differences. However, these methods usually shrinks both precision matrices and the difference of precision matrices. Their performance is thus limited if precision matrices are not sparse. One exception is the fused graphical lasso method proposed by \citet{danaher2014joint}. The fused graphical lasso method does not shrink precision matrices (if the penalty for the precision matrix is set as zero), but there was no statistical theory that guarantees its consistency. More recently, \citet{zhao2014direct} extended their $l_1$-minimization method for sparse precision matrix estimation \citep{cai2011constrained} and developed a new $l_1$-minimization method for differential network analysis. The authors proved asymptotic results without assuming sparsity of precision matrices. However, both of the computational complexity and the memory requirement of the $l_1$-minimization method are around $O(p^4)$. When $p$ is relatively large, it will be computationally prohibitive to calculate. A few other researchers also considered the differential network analysis \citep{li2007finding,mohan2012structured,zhang2012learning}, but there was no theoretical result developed for these methods.

In this paper, we propose a new smooth and convex loss function to directly estimate the precision matrix difference, without attempting to estimate the precision matrices individually. This loss function can be viewed as a generalization of the D-trace loss in \citet{zhang2014sparse} and hence we also call it the D-trace loss. By adding a lasso penalty to this D-trace loss, we can estimate the precision matrix difference in high-dimensional setting. This D-trace loss function takes a very simple form and hence allows us derive consistency theory for sub-Gaussian as well as polynomial-tailed distributions under a new irrepresentability condition. We show that the irrepresentability condition is less stringent than the mutual incoherence condition used in \citet{zhao2014direct}. The simplicity of the D-trace loss function also allows us to develop an efficient algorithm. Simulation studies and a real data analysis showed that this lasso penalized D-trace loss estimator outperforms other available methods. The paper is organized as following. We will introduce the D-trace loss function, present the algorithm for solving the lasso penalized D-trace loss function in Section 2. Section 3 discusses the consistency results. Simulation Studies and a real data analysis are presented in Section 4 and Section 5, respectively. Section 6 presents discussions of extensions and future research directions.

\section{Methods}
\subsection{The D-trace Loss Function}
\label{sec2}
Suppose that $A=(A_{i,j})\in \mathbb{R}^{p\times p}$ is a $p\times p$ matrix, we denote $ \|A\|_F=(\sum_{i,j}A_{i,j}^2)^{1/2}$ as its Frobenius norm and $vec(A)$ as the $p^2$-vector by stacking the columns of X. Let $<A,B>=tr(AB^T)$ and we have $<A,A> = \|A\|_F^2$. Our goal is to find a matrix $\Delta$ to estimate $\Sigma_Y^{-1}-\Sigma_X^{-1}$. To do this, we first construct a new convex loss function $L(\Delta, \Sigma_X, \Sigma_Y)$ such that its unique minimizer given $\Sigma_X$ and $\Sigma_Y$ is achieved at $\Delta=\Sigma_Y^{-1}-\Sigma_X^{-1}$. In other words, the minimizer of the loss function $L(\Delta, \Sigma_X, \Sigma_Y)$ should satisfy $\Sigma_X\Delta\Sigma_Y - (\Sigma_X - \Sigma_Y) = 0$ and $\Sigma_Y\Delta\Sigma_X - (\Sigma_X - \Sigma_Y) =0$, and thus $(\Sigma_X\Delta\Sigma_Y + \Sigma_Y\Delta\Sigma_X)/2 - (\Sigma_X - \Sigma_Y) = 0$. If we define the loss function $L_D(\Delta, \Sigma_X, \Sigma_Y)$ as the following D-trace loss function
\begin{equation}
\label{LDestim}
L_D(\Delta, \Sigma_X, \Sigma_Y)=\frac{1}{4}(<\Sigma_X\Delta,\Delta\Sigma_Y>+<\Sigma_Y\Delta,\Delta\Sigma_X>)-<\Delta, \Sigma_X-\Sigma_Y>,
\end{equation}
we have
\begin{equation}
\label{LD_Derivative}
\frac{\partial L_D}{\partial \Delta} = (\Sigma_X\Delta\Sigma_Y + \Sigma_Y\Delta\Sigma_X)/2 - (\Sigma_X - \Sigma_Y).
\end{equation}
It is easy to check that the Hessian matrix with respect to $\Delta$ of the D-trace loss function (\ref{LDestim}) is $(\Sigma_X\bigotimes\Sigma_Y+\Sigma_Y\bigotimes\Sigma_X)/2$, where $\bigotimes$ is the Kronecker product. Therefore, the loss function $L_D$ is a convex function about $\Delta$ and has a unique minimizer at $\Delta=\Sigma_Y^{-1}-\Sigma_X^{-1}$. Suppose that $\hat{\Sigma}_X$ and $\hat{\Sigma}_Y$ are sample covariance matrices of $X_i$ ($i=1,\cdots,n_X$) and $Y_j$ ($j=1,\cdots,n_Y$), respectively. With the loss function $L_D$, we can estimate $\Delta$ by minimizing the following lasso penalized loss function,
\begin{equation}
\label{LDpenal}
L_D(\Delta, \hat{\Sigma}_X, \hat{\Sigma}_Y)+\lambda\|\Delta\|_{1},
\end{equation}
where $\lambda>0$ is a tuning parameter. We develop an efficient alternating method (AMD) for minimizing the objective function (\ref{LDpenal}) in Section \ref{subsection:algorithm}. Theoretical results are developed in Section \ref{s:inf}.

\subsection{Algorithm}
\label{subsection:algorithm}
Directly minimizing the objective function (\ref{LDestim}) is difficult, we first introduce two auxiliary matrices $\Delta_1$ and $\Delta_2$ and consider the following minimization problem
\begin{equation}
\label{alter}
\begin{aligned}
\min_{\Delta=\Delta_1=\Delta_2} &L_1(\Delta_1,\hat\Sigma_X,\hat\Sigma_Y) + L_2(\Delta_2,\hat\Sigma_X,\hat\Sigma_Y)+\lambda \|\Delta_3\|_{1},
\end{aligned}
\end{equation}
where $4L_1(\Delta,\Sigma_X,\Sigma_Y)=<\Sigma_X\Delta,\Delta\Sigma_Y> - 2<\Delta, \Sigma_X-\Sigma_Y>$ and $4L_2(\Delta,\Sigma_X,\Sigma_Y) = <\Sigma_Y\Delta,\Delta\Sigma_X> - 2<\Delta, \Sigma_X-\Sigma_Y>$. Note that solving (\ref{alter}) is equivalent to minimizing (\ref{LDpenal}) since $L_D(\Delta,\Sigma_X,\Sigma_Y)=L_1(\Delta,\Sigma_X,\Sigma_Y)+L_2(\Delta,\Sigma_X,\Sigma_Y)$. With (\ref{alter}), we consider the augmented Lagrangian
\begin{equation*}
\begin{aligned}
L(\Delta_1, \Delta_2, \Delta_3, \Lambda_1, \Lambda_2, \Lambda_3)&=L_1(\Delta_1,\hat\Sigma_X,\hat\Sigma_Y) + L_2(\Delta_2,\hat\Sigma_X,\hat\Sigma_Y)+\lambda \|\Delta_3\|_{1}\\
&+<\Lambda_1, \Delta_3-\Delta_1>+<\Lambda_2, \Delta_2-\Delta_3>+<\Lambda_3, \Delta_1-\Delta_2>\\
&+(\rho/2)\|\Delta_3-\Delta_1\|_F^2+(\rho/2)\|\Delta_2-\Delta_3\|_F^2+(\rho/2)\|\Delta_1-\Delta_2\|_F^2.
\end{aligned}
\end{equation*}
In our ADM algorithm, we choose $\rho>0$ to be a fixed number and iteratively update $\Delta_1$,$\Delta_2$,$\Delta_3$, $\Lambda_1$,$\Lambda_2$,$\Lambda_3$. Specifically, given $\Delta_1^k$, $\Delta_2^k$, $\Delta_3^k$, $\Lambda_1^k$,$ \Lambda_2^k$ and $\Lambda_3^k$ at the $k$th step , we update the estimates as the following
\begin{eqnarray}
\label{Delta}
\Delta_1^{k+1}&=&\mbox{argmin}_{\Delta_1}L(\Delta_1,\Delta_2^k, \Delta_3^k,\Lambda_1^k,\Lambda_2^k,\Lambda_3^k)\\
\label{Delta1}
\Delta_2^{k+1}&=&\mbox{argmin}_{\Delta_2}L(\Delta_1^{k+1}, \Delta_2,\Delta_3^{k},\Lambda_1^k,\Lambda_2^k,\Lambda_3^k)\\
\label{Delta2}
\Delta_3^{k+1}&=&\mbox{argmin}_{\Delta_3}L(\Delta_1^{k+1}, \Delta_2^{k+1},\Delta_3,\Lambda_1^k,\Lambda_2^k,\Lambda_3^k)\\
\nonumber
\label{lambda}
\Lambda_1^{k+1}&=&\Lambda_1^k+\rho(\Delta_3^{k+1}-\Delta_1^{k+1})\\
\nonumber
\label{lambda1}
\Lambda_2^{k+1}&=&\Lambda_2^k+\rho(\Delta_2^{k+1}-\Delta_3^{k+1})\\
\nonumber
\label{lambda2}
\Lambda_3^{k+1}&=&\Lambda_2^k+\rho(\Delta_1^{k+1}-\Delta_2^{k+1})
\nonumber
\end{eqnarray}
For (\ref{Delta}), if we take partial derivative about $\Delta_1$ of the objective function and setting it as zero, we get
\begin{eqnarray*}
\hat\Sigma_X\Delta_1\hat\Sigma_Y/2 + 2\rho \Delta_1 - \rho(\Delta_3^k+\Delta_2^k) - (\hat\Sigma_X-\hat\Sigma_Y)/2 - \Lambda_1^k + \Lambda_3^k = 0.
\end{eqnarray*}
Thus, $\Delta_1^{k+1}$ (and similarly $\Delta_2^{k+1}$) satisfies equation of the form
\begin{equation}
\label{eqnUpdateDelta}
AXB+\gamma X=C,
\end{equation} where $A$ and $B$ are symmetric, nonnegative definite matrices, $\gamma>0$ is a constant and $C$ is a matrix. Explicit solution to the equation (\ref{eqnUpdateDelta}) is given in the following Lemma. The proof of this lemma is given in the Supplementary.
\begin{lemma}
\label{lemmaG}
Let $A=U_A\Sigma_AU_A^T$ and $B=U_B\Sigma_BU_B^T$ be the eigenvalue decompositions of the symmetric matrices $A$ and $B$, respectively. Assume that $G(A, B, C, \rho)$ is the solution to (\ref{eqnUpdateDelta}). Then,  $G(A,B,C,\gamma)=U_A[D\circ(U_A^TCU_B^T)]U_B$, where $D_{ij}=(\sigma_j^A\sigma_i^B+\gamma)^{-1}$ and $\circ$ denotes the Hadamard product of two matrices.
\end{lemma}

Given a matrix $A$ and $\lambda>0$, let $S(A,\lambda)$ be the solution to the following optimization problem
\begin{equation*}
S(A,\lambda) = \mbox{argmin}_{\Delta}\frac{1}{2}\|\Delta\|_F^2-<\Delta,A>+\lambda\|\Delta\|_{1}.
\end{equation*}
It is easy to check that the $(i,j)$th component of $S(A,\lambda)$ is
\begin{numcases}{S(A,\lambda)_{i,j}=}
  A_{i,j}-\lambda & $A_{i,j}>\lambda$, \notag\\
  A_{i,j}+\lambda & $A_{i,j}<-\lambda$, \notag\\
  0 & $-\lambda\leq A_{i,j}\leq\lambda$. \notag
\end{numcases}
The optimization problem (\ref{Delta2}) is equivalent to
\begin{equation*}
\mbox{argmin}_{\Delta_3}\rho\|\Delta_3\|_F^2- <\Delta_3,\rho\Delta_1^{k+1} + \rho\Delta_2^{k+1}-\Lambda_1^k+\Lambda_2^k> + \lambda \|\Delta_3\|_{1},
\end{equation*}
and thus $\Delta_3^{k+1} = S\left((\rho\Delta_1^{k+1} + \rho\Delta_2^{k+1}-\Lambda_1^{k}+\Lambda_2^{k})/2\rho,\lambda/2\rho\right)$. We summarize the AMD algorithm in Algorithm \ref{alg:Framwork}. In our simulation and real data analysis, we let $\rho=50$ and terminate the algorithm if ${\|\Delta_j^{k+1}-\Delta_j^k\|_F}<10^{-3}\max(1,\|\Delta_j^k\|_F,\|\Delta_j^{k+1}\|_F)$ ($j=1,2,3$).
\begin{algorithm}[htb]
\caption{The AMD algorithm for the lasso penalized D-trace loss estimator.} 
\label{alg:Framwork} 
 \vspace*{-12pt}
 \begin{tabbing}
\enspace Initialization: k=0, $\Delta_1^0,\Delta_2^0,\Delta_3^0=(diag(\hat{\Sigma}_Y)+I)^{-1}-(diag(\hat{\Sigma}_X)+I)^{-1}$,$\Lambda^0,\Lambda_1^0,\Lambda_2^0=0$.\\
\enspace Given $\Delta_3^k$, $\Delta_1^k$, $\Delta_2^k$, $\Lambda_1^k$,$ \Lambda_2^k$ and $\Lambda_3^k$ at the $k$th step, at the $k+1$th step, we update\\
\qquad (a) $\Delta_1^{k+1}=G\left(\hat\Sigma_X,\hat\Sigma_Y,2\rho\Delta_3^k+2\rho\Delta_2^k+  \hat\Sigma_X-\hat\Sigma_Y+2\Lambda_1^k-2\Lambda_3^k,4\rho\right)$\\
\qquad (b) $\Delta_2^{k+1}=G\left(\hat\Sigma_Y,\hat\Sigma_X,2\rho\Delta_3^k+2\rho\Delta_1^{k+1}+ \hat\Sigma_X-\hat\Sigma_Y+2\Lambda_3^k-2\Lambda_2^k,4\rho\right)$\\
\qquad (c) $\Delta_3^{k+1} = S\left(\frac{1}{2\rho}(\rho\Delta_1^{k+1} + \rho\Delta_2^{k+1}-\Lambda_1^{k}+\Lambda_2^{k}),\frac{\lambda}{2\rho}\right)$\\
\qquad (d) $\Lambda_1^{k+1}=\Lambda_1^k+\rho(\Delta_3^{k+1}-\Delta_1^{k+1})$\\
\qquad (e) $\Lambda_2^{k+1}=\Lambda_2^k+\rho(\Delta_2^{k+1}-\Delta_3^{k+1})$,\\
\qquad (f) $\Lambda_3^{k+1}=\Lambda_2^k+\rho(\Delta_1^{k+1}-\Delta_2^{k+1})$.\\
\enspace Repeat steps (a-f) until convergence.\\
\enspace Output $\Delta_3^{k+1}$ as the estimate of the difference of the precision matrices $\Delta^*$.
 \end{tabbing}
\end{algorithm}

It is easy to see that the computational complexity of each iteration in Algorithm \ref{alg:Framwork} is $O(p^3)$. Since we only need to store a few matrices in the memory, the memory requirement of the above algorithm is only $O(p^2)$. In comparison, the computational complexity and the memory requirement of the $l_1$-minimization algorithm \citep{zhao2014direct} are all $O(p^4)$. Lastly, we select the tuning parameter by minimizing the Bayesian Information Criterion (BIC). For our method, the BIC is defined as
\begin{equation*}
(n_X+n_Y)\|\frac{1}{2}(\hat{\Sigma}_X\Delta\hat{\Sigma}_Y+\hat{\Sigma}_Y\Delta\hat{\Sigma}_X)-\hat{\Sigma}_X+\hat{\Sigma}_Y\|+\log(n_X+n_Y)|\Delta|_0,
\end{equation*}
where the norm $\|\cdot\|$ can be the $L_F$-norm or the $L_\infty$-norm, and $|\Delta|_0$ denotes the number of non-zero elements in $\Delta$.
For other two methods, following \citet{zhao2014direct}, the BIC is defined as
$(n_X+n_Y)\|\hat{\Sigma}_X\Delta\hat{\Sigma}_Y-\hat{\Sigma}_X+\hat{\Sigma}_Y\|+\log(n_X+n_Y)|\Delta|_0.$

\section{Theoretical Properties}
\label{s:inf}
In this section, we study the theoretical properties of the proposed estimator in ultra-high dimensional setting.

\subsection{The irrepresentability condition}
 We assume that the true network difference $\Delta^*$ is sparse, $S=\{(i,j): \Delta_{i,j}^*\neq 0\} $ is the support of $\Delta^*$ and $s=\mid S\mid$. Given a vector $v \in \mathcal{R}^n$, we use $\|v\|_1$ and $\|v\|_\infty$ as its $L_1$-norm and $L_\infty$-norm, respectively. Given a matrix $A$, we denote $\|A\|_1=\|\mbox{vec}(A)\|_1$, $\|A\|_\infty=\|\mbox{vec}(A)\|_\infty$. In addition, we define $\|A\|_{1,\infty}=\max_i\sum_j\mid A_{i,j}\mid$ as the $L_{1,\infty}$ norm of matrix $A$. Suppose that $\Gamma=A\otimes B$, where $A = (A_{j,l})$ and $B=(B_{k,m})$ are two $p\times p$ matrices. For any two subsets $T_1$ and $T_2$ of $\{1,\ldots,p\}\times\{1,\ldots,p\}$, we denote by $\Gamma_{T_1T_2}$ the submatrix of $\Gamma$ with rows and columns indexed by $T_1$ and $T_2$, i.e., we have $\Gamma_{T_1T_2} = \big(A_{j,l}B_{k,m}\big)_{(j,k)\in T_1,~(l,m)\in T_2}$.

The theoretical properties discussed in this section will be based on a new irrepresentability condition. Denote
$\Gamma(\Sigma_X,\Sigma_Y)=(\Sigma_X \otimes \Sigma_Y+\Sigma_Y \otimes \Sigma_X)/2.$
For notation simplicity, we write $\Gamma^*=\Gamma(\Sigma_X^*,\Sigma_Y^*)=(\Gamma_{ij}^*)$ and $\hat{\Gamma}=\Gamma(\hat{\Sigma}_X,\hat{\Sigma}_Y)$. We assume the following irrepresentability condition
\begin{equation}
\label{Irre}
\max_{e\in S^c}{\|\Gamma_{e,S}^* {(\Gamma_{S,S}^*)}^{-1}\|}_1<1.
\end{equation}
Suppose that $\alpha=1-\max_{e \in S^c}\|\Gamma_{e,S}^*(\Gamma_{S,S}^*)^{-1}\|_1$ and $\kappa_\Gamma=\|\Gamma_{S,S}^{*-1}\|_{1,\infty}$. Then, we have $\alpha>0$. The irrepresentability condition (\ref{Irre}) takes a very similar form as the ones used in \citet{5354541} and \citet{zhang2014sparse}. For any $(j,k)$, let $Z_{(j,k)}=X_jY_k$. We have $(\Sigma_X \otimes \Sigma_Y)_{(j,l),(k,m)} = E(Z_{(j,k)}Z_{(l,m)})$. Thus, roughly speaking, the irrepresentability condition (\ref{Irre}) enforces that the edge variable $Z_{(j,k)}$ not in the difference network ($(j,k)\in S^c$) and the edge variable $Z_{(l,m)}$ in the difference network ($(l,m)\in S$) cannot be highly correlated.

It is interesting to compare the condition (\ref{Irre}) with Condition 2 in \citet{zhao2014direct}. Define $\Sigma_{\max}^X=\max_j\Sigma_{X,jj}^*$, $\Sigma_{\max}^Y=\max_j\Sigma_{Y,jj}^*$, $\mu_X=\max_{i\neq j} |\Sigma_{X,ij}^*|$ and $\mu_Y=\max_{i\neq j} |\Sigma_{Y,ij}^*|$. Condition 2 implies that $\mu=\max(\mu_Y\Sigma_{\max}^X, \mu_X\Sigma_{\max}^Y)\leq \min_{j,k}(\Sigma_{X,jj}^*\Sigma_{Y,kk}^*)(2s)^{-1}$, which in turn implies that $\max_{i\neq j}|\Gamma^*_{ij}| \leq \min_{j,k}(\Sigma_{X,jj}^*\Sigma_{Y,kk}^*)(2s)^{-1}$. Since  $\min_j\Gamma^*_{jj} = \min_{j,k}(\Sigma_{X,jj}^*\Sigma_{Y,kk}^*)$, we can prove the irrepresentability condition (\ref{Irre}) using the similar technique as in the proof of Corollary 2 of \citet{zhao2006model}. Therefore, Condition 2 in \citet{zhao2014direct} is a stronger condition than the irrepresentability condition (\ref{Irre}).

We give an example that satisfies the irrepresentability condition but not Condition 2 in \citet{zhao2014direct}. Suppose that $\Sigma_X^*=\mbox{diag}\{A,B_X\}$ and $\Sigma_Y^*=\mbox{diag}\{A,B_Y\}$, where $A$ is a symmetric positive definite $p_1\times p_1$ matrix and $B_X$ and $B_Y$ are symmetric positive definite $p_2\times p_2$ matrix matrices ($p_1+p_2=p$, $p_2\geq 1$). We can take $A$, $B_X$, $B_Y$ such that the maximum diagonal term of $\Sigma_X^*$ and $\Sigma_Y^*$ is 1 and the maximum absolute off-diagonal term is $1>\rho>1/2$. Take $s=p_2^2<p$ and the matrices $B_X$ and $B_Y$ such that the corresponding elements of $B_X^{-1}$ and $B_Y^{-1}$ are all different. Thus, we have $\mu> 1/2$ and $\min_{j,k}(\Sigma_{X,jj}^*\Sigma_{Y,kk}^*)(2s)^{-1} < 1/2$. Therefore, Condition 2 in Zhao et al. 2014 does not hold. On the other hand, with this choice of $\Sigma_X^*$ and $\Sigma_Y^*$, we have $S=\{(i,j)|~p_1 < i,j\leq n\}$ and it can be easily verified that $\max_{e\in S^c}{\|\Gamma_{e,S}^* {(\Gamma_{S,S}^*)}^{-1}\|}_1=0< 1$ and thus the irrepresentability condition holds. In addition, Condition 1 in \citet{zhao2014direct} requires $|\Delta^*|_1$ is bounded. This is a relatively strong condition, because if we assume the nonzero elements of $\Delta^*$ is bounded away from zero, boundedness of $|\Delta^*|_1$ would imply that $s$ is bounded.

\subsection{Convergence Rates}
We introduce some notations before giving the theoretical results. Recall that a mean-zero random vector $Z\in\mathcal{R}^p$ with covariance matrix $\Sigma$ is called sub-Gaussian if there exists a constant $\sigma\in (0,\infty)$ such that $\mbox{E}[\exp\{tZ_i(\Sigma_{ii})^{-1/2}\}]\leq \exp(\sigma^2t^2/2)$ for all $t\in \mathcal{R}$ and $i=1,\cdots,p$, where $\Sigma_{ii}$ is the $(i,i)$th element of $\Sigma$. It is called having a polynomial tail if there exists a positive integer $m$ and scalar $K_m\in\mathcal{R}$ such that $\mbox{E}[\exp\{tZ_i(\Sigma_{ii})^{-1/2}\}]^{4m}\leq K_m$ \citep{5354541}. Given random vectors $X_i$ and $Y_j$, we assume that they are independent and $X$s ($Y$s) have the same distribution ($X$ and $Y$ generally have different distributions). We always assume $s<p$ and $\max\{\|\Sigma_X^*\|_\infty,\|\Sigma_Y^*\|_\infty\}\leq M$ for some constant $M$ independent of $p$. If they are sub-Gaussian distributions, we assume their associated constants are $\sigma_X$ and $\sigma_Y$, respectively. If they are of polynomial-tail, we assume their associated constants are $K_{Xm}$ and $K_{Ym}$. To state the theorems, we define the following notations,
\begin{equation*}
\begin{aligned}
&\tilde{M}={24sM(2sM^2\kappa_\Gamma^2+\kappa_\Gamma)}/{\alpha},\\
&\tilde{\delta}_{GZ}=\max_i(\Sigma_{Z,ii}^*)^2(1+4\sigma_Z^2)^2,~~\tilde{\delta}_{PZ}=\max_i\Sigma_{Z,ii}^*(1+K_{Zm})^{1/(2m)}~~Z\in\{X,Y\},\\
&G_A=\frac{{\tilde{\delta}_{GX}}^{1/2}}{{n_X}^{1/2}}+\frac{{\tilde{\delta}_{GY}}^{1/2}}{{n_Y}^{1/2}},~~G_B=\frac{{\tilde{\delta}_{GX}}^{1/2}}{{n_X}^{1/2}}\frac{{\tilde{\delta}_{GY}}^{1/2}}{{n_Y}^{1/2}},P_A=\frac{\tilde{\delta}_{PX}}{{n_X}^{1/2}}+\frac{\tilde{\delta}_{PY}}{{n_Y}^{1/2}},~~P_B=\frac{\tilde{\delta}_{PX}}{{n_X}^{1/2}}\frac{\tilde{\delta}_{PY}}{{n_Y}^{1/2}}.\\
\end{aligned}
\end{equation*}
We first establish the theoretical properties for the sub-Gaussian distributions.
\begin{theorem}
\label{theorem1}
Assume that $X_i$, $Y_j$ are sub-Gaussian with parameter $\sigma_X$ and $\sigma_Y$, respectively. Under the irrepresentability condition (\ref{Irre}), if
\begin{equation*}
\begin{aligned}
\lambda_n &=\max\left[{2(4-\alpha)}G_A/\alpha, \{128(\eta\log p+\log 4)\}^{1/2}\tilde{M}G_B+MG_A\tilde{M}\right]\{128(\eta\log p+\log 4)\}^{1/2}\\
 \end{aligned}
\end{equation*}for some $\eta>2$ and $\min(n_X,n_Y)>C_{G}\bar{\delta}^{-2}(\eta \log p+\log 4)$, then, with probability larger than $1-2/p^{\eta-2}$, we have that the support of $\hat{\Delta}$ is in the support of $\Delta^*$ and that
\begin{equation*}
\begin{aligned}
 \|\hat{\Delta}-\Delta^*\|_\infty \leq M_G\bigg\{\frac{\eta \log p+\log 4}{\min{(n_X, n_Y)}}\bigg\}^{1/2},~~ \|\hat{\Delta}-\Delta^*\|_F\leq M_G\bigg\{\frac{\eta \log p+\log 4}{\min{(n_X, n_Y)}}\bigg\}^{1/2}s^{1/2},\\
 \end{aligned}
\end{equation*}
where $\bar{\delta}$, $C_{G}$, $M_{G}$ are constants depending on $M$, $s$, $\kappa_\Gamma$, $\alpha$, $\sigma_X$ and $\sigma_Y$ (See Appendix for their definitions).
\end{theorem}
Let $(j,k)th$ entry of $\hat{\Delta}$ be $\hat{\Delta}_{j,k}$, and $sgn(t)$ be the sign function.
Denote $M(\hat{\Delta})=\{sgn(\hat{\Delta}_{j,k}): j=1, \dots, p, k=1, \ldots, p\}$ and $M(\Delta^*)=\{sgn(\Delta^*_{j,k}): j=1, \dots, p, k=1, \ldots, p\}$. We can have the following sign consistency result from Theorem \ref{theorem1}.
\begin{theorem}
\label{theorem2}
Under the same conditions and notations in Theorem \ref{theorem1}, if
\begin{equation*}
\min_{j,k: \Delta_{j,k}^*\neq 0}{\mid \Delta_{j,k}^*\mid }\geq 2M_G~\bigg\{\frac{\eta \log p+\log 4}{\min{(n_X, n_Y)}}\bigg\}^{1/2}
\end{equation*}
 for some $\eta>2$ and, then $M(\hat{\Delta})=M(\Delta^*)$ with probability $1-2/p^{\eta-2}$.
\end{theorem}

For random vectors with polynomial tails, we also have the following results about the rates of convergence and the model selection consistency.
\begin{theorem}
\label{theorem3}
Assume that $X_i$, $Y_j$ are of polynomial tail with parameters $(m,K_{Xm})$ and $(m,K_{Ym})$, respectively. Under the irrepresentability condition (\ref{Irre}), take
\begin{equation*}
\begin{aligned}
\lambda_n&=\max\{{2(4-\alpha)}P_A/\alpha, \tilde{M}(2p^{\eta/(2m)}P_B+MP_A)\}2p^{\eta/(2m)}
\end{aligned}
\end{equation*} for some $\eta>2$ and $\min(n_X,n_Y)>C_{P}\bar{\delta}^{-2}p^{\eta/m},$
then with probability larger than $1-2/p^{\eta-2}$, we have that the support of $\hat{\Delta}$ is in the support of $\Delta^*$ and that
\begin{equation*}
\begin{aligned}
 \|\hat{\Delta}-\Delta^*\|_\infty  \leq \frac{M_P~p^{\eta/(2m)}}{\min{(n_X, n_Y)}^{1/2}},~~ \|\hat{\Delta}-\Delta^*\|_F \leq \frac{M_P~p^{\eta/(2m)}s^{1/2}}{\min{(n_X, n_Y)}^{1/2}},\\
 \end{aligned}
\end{equation*}
where $\bar{\delta}$, $C_P$, $M_P$ are constants depending on $M$, $s$, $\kappa_\Gamma$, $\alpha$, $K_{m_X}$, $K_{m_Y}$, and $m$ (see Appendix).
\end{theorem}

\begin{theorem}
\label{theorem4}
Under the same conditions and notations in Theroem \ref{theorem3}, if
\begin{equation*}
\min_{j,k: \Delta_{j,k}^*\neq 0}{\mid \Delta_{j,k}^*\mid}\geq 2M_P~p^{\eta/(2m)}\min{(n_X, n_Y)}^{-1/2}
\end{equation*}
 for some $\eta>2$, then $M(\hat{\Delta})=M(\Delta^*)$ with probability $1-2/p^{\eta-2}$.
\end{theorem}

The techniques for proving Theorems 1-4 are similar to the proofs used in \citet{zhang2014sparse}, although the proofs here are more complicated because there are two covariance matrices involved. The error bounds we obtained are exactly in parallel to those in \citet{zhang2014sparse} and \citet{5354541}. For example, similar to Theorem \ref{theorem1}, \citet{zhang2014sparse} showed that the error bound of their precision matrix estimation for Gaussian data under $L_\infty$-norm is $\tilde{M}_G\left\{(\eta\log p + \log 4)/n\right\}^{1/2}$, where $\tilde{M}_G$ depends on constants similar to $\kappa_\Gamma$ and $\alpha$ and $n$ is the number of observations. \citet{zhao2014direct} showed that their estimator of the precision matrix difference $\tilde{\Delta}$ satisfies
$\|\tilde{\Delta}-\Delta^*\|_\infty \leq M_Z\left\{\log p/{\min{(n_X, n_Y)}}\right\}^{1/2}$, where $M_Z$ depends on a number of characteristics of $\Sigma_X^*$ and $\Sigma_Y^*$. It is difficult to directly compare this error bound with the error bound in Theorem \ref{theorem1}. However, if $s$, $\Sigma_X^*$ and $\Sigma_Y^*$ are bounded, $\sigma_{\min}^S$ as defined in \citet{zhao2014direct} is bounded away from zero and all conditions in \citet{zhao2014direct} holds, we can easily show that $\hat{\Delta}$ and $\tilde{\Delta}$ converges at the same rate of $O(\log p/\min(n_X,n_Y))$ for Gaussian data under $L_\infty$-norm.

\section{Simulation Studies}
\label{sec3}
In this section, we perform simulations to compare the performance of our D-trace loss estimator with the fused graphical lasso method \citep{danaher2014joint} and the $l_1$-minimization method \citep{zhao2014direct}. Across all simulation setups, we set $p=100,200,500$ and $1000$ and $n_X=n_Y=100,200,500$. Each simulation was repeated 100 times. We generated $X_i$s and $Y_i$s from normal distributions. The covariance matrices $\Sigma_X$ and $\Sigma_Y$ were generated differently in different simulation setups. For the $l_1$-minimization method, we only performed the simulation for $p=100$ because it is computationally too expensive for $p=200$, $500$ and $1000$.

\begin{figure}[!htpb]
\centering
\subfigure[]{
\includegraphics[angle=270,width=6cm]{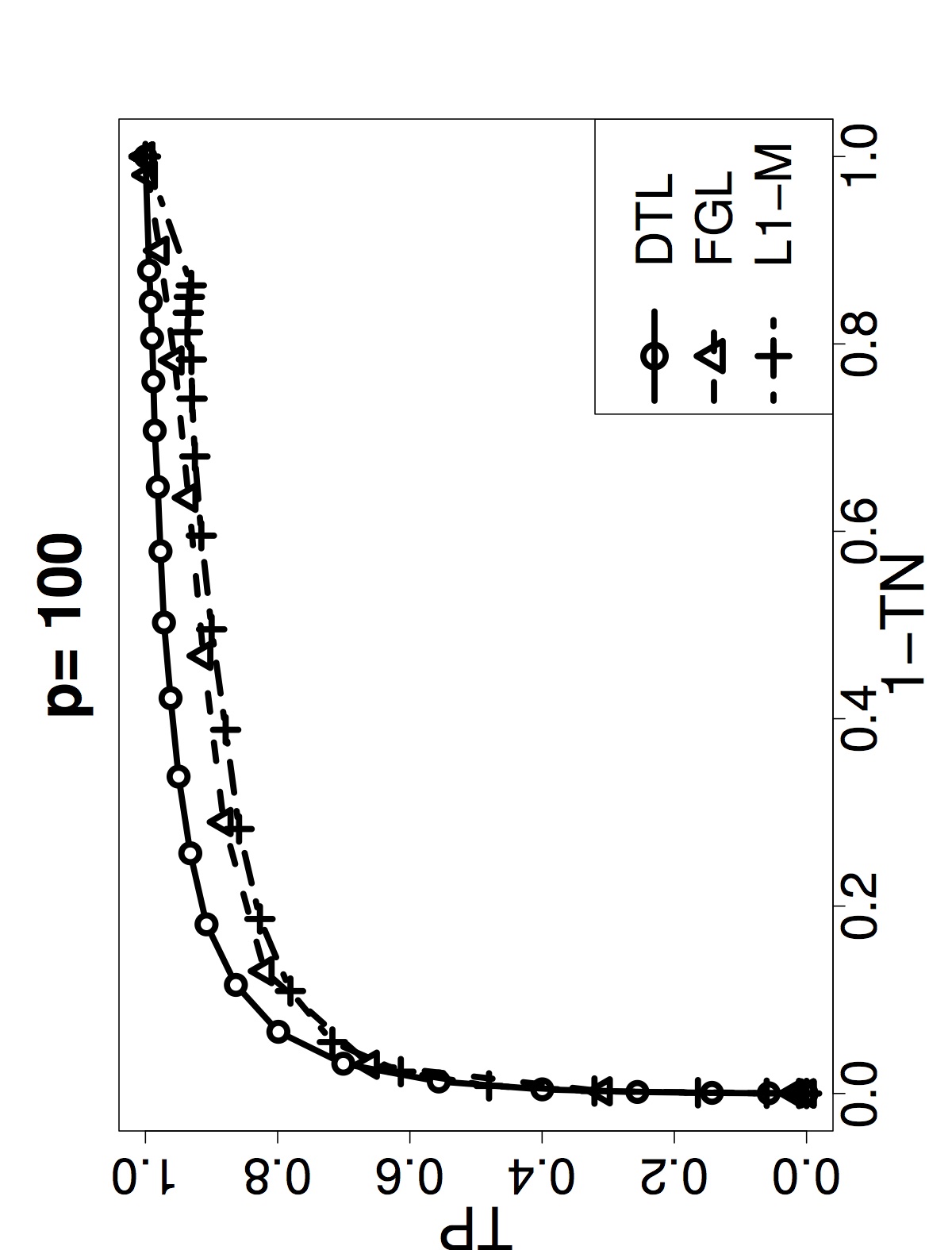}}
\subfigure[]{
\includegraphics[angle=270,width=6cm]{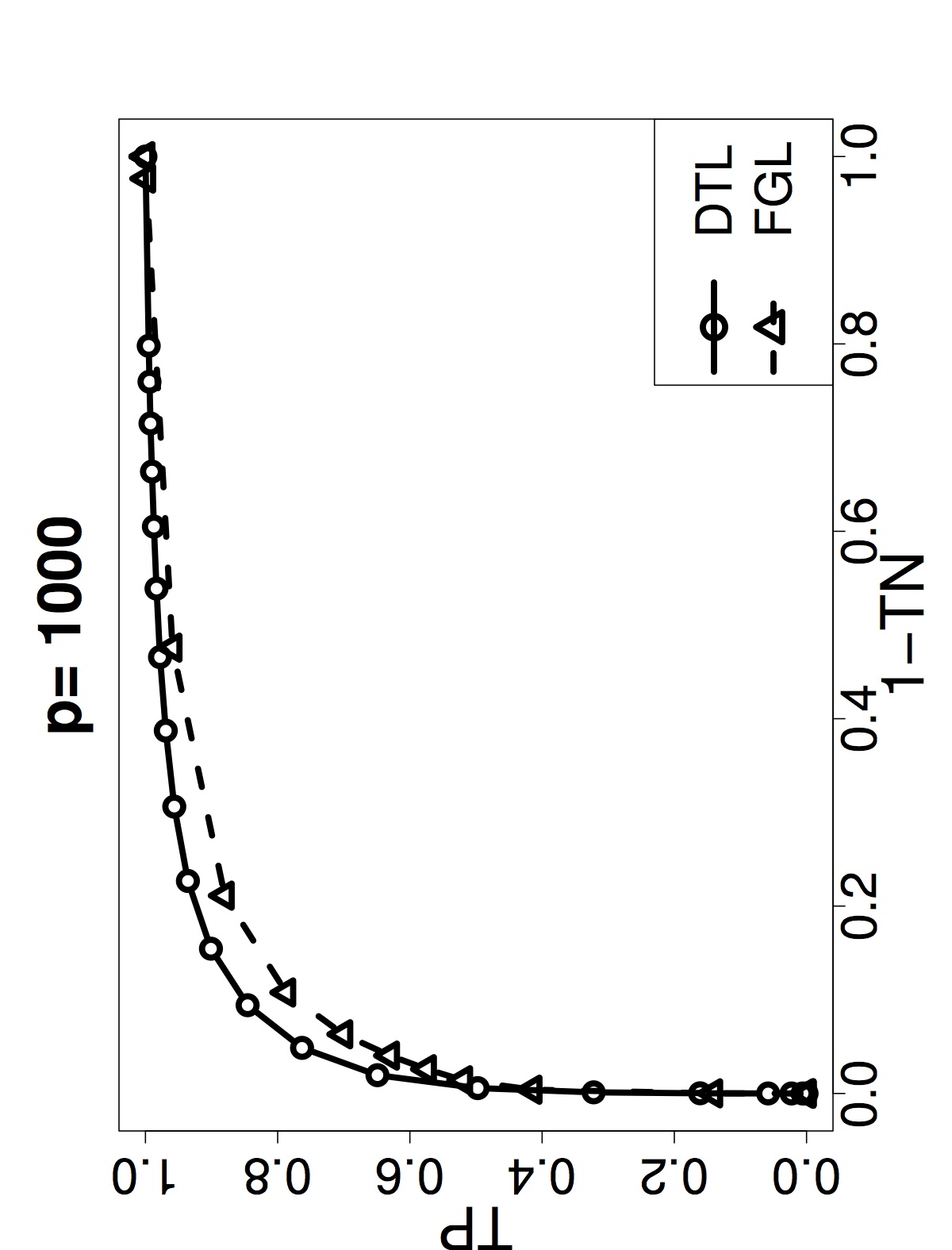}}
\subfigure[]{
\includegraphics[angle=270,width=6cm]{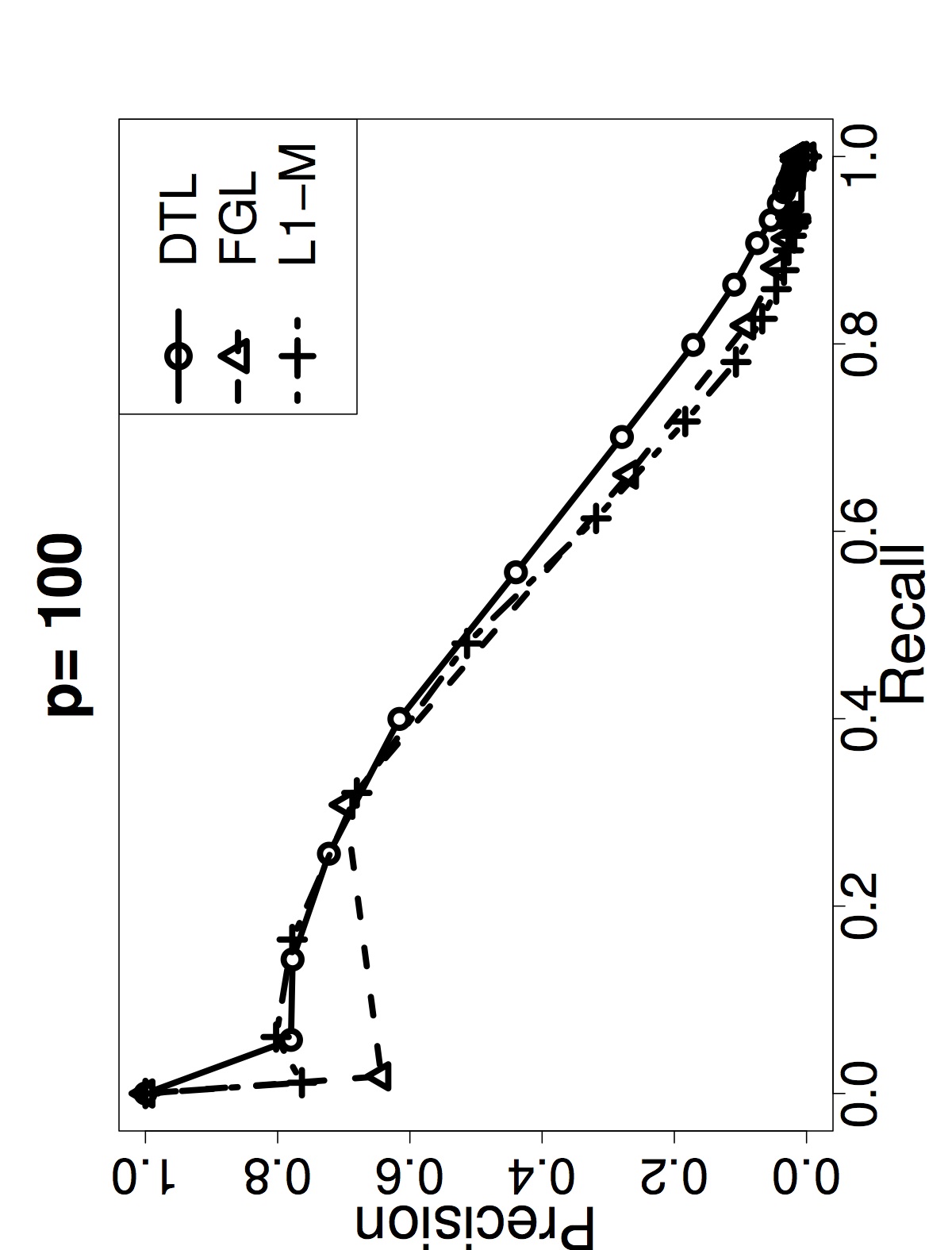}}
\subfigure[]{
\includegraphics[angle=270,width=6cm]{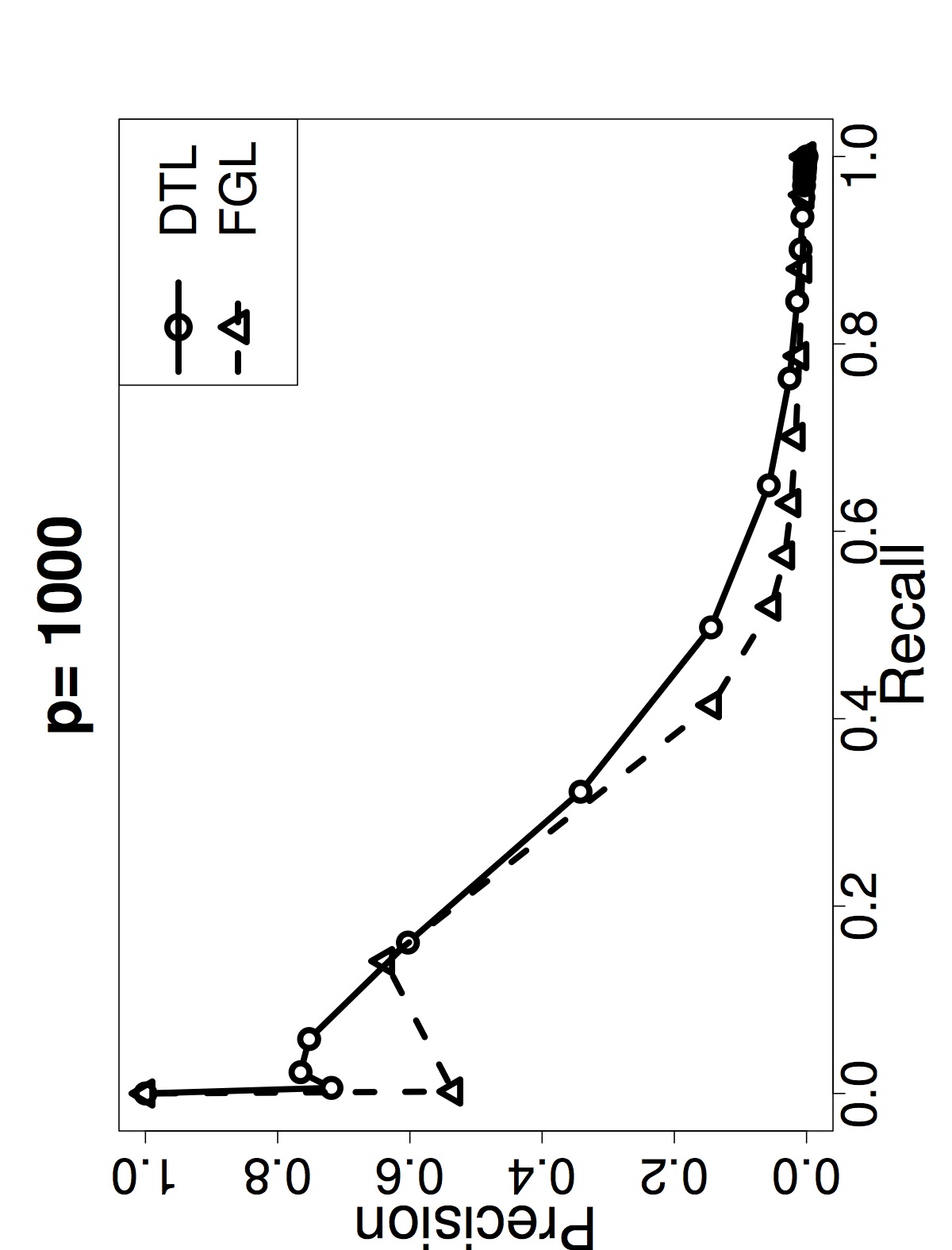}}
\caption{The ROC curve (a,b) and the Precision-Recall curve (c-d) for support recovery of $\Delta^*$ with $p=100$ and $1000$ for Simulation 1 when n=100. In the figure, DTL stands for the D-trace loss, FGL for the fused graphical lasso and L1-M for the $l_1$-minimization method in \citet{zhao2014direct}. }\label{Fig:simulation_ROC}
\end{figure}

\begin{itemize}
\item Simulation 1.In this simulation, $(i ,j)$ element in the precision matrix $\Sigma_X^{-1}$ was defined as $0.5^{|i-j|}$, and the precision matrix $\Sigma_Y^{-1}$ was similar except that elements which satisfy $|i-j|=\lfloor p/4\rfloor$ were defined as 0.9, where $\lfloor x\rfloor$ means taking integer part of $x$.

\item Simulation 2. The precision matrices had block structures. Each block was a $50\times 50$ matrix, and there were two blocks when p=100, four blocks when p=200, ten blocks when p=500 and twenty blocks when p=1000. In each block, the precision matrices $\Sigma_X^{-1}$ and $\Sigma_Y^{-1}$ were generated in the same way as in \citet{zhao2014direct}. Briefly, the support of $\Sigma_X^{-1}$ was first generated according to a network with $50\times(50-1)/10$ edges and a power law degree distribution with an expected power parameter of 2. A uniform distribution with support $[-0.5,-0.2]\cup[0.2,0.5]$ was used to generate the nonzero entry of $\Sigma_X^{-1}$. Each row of $\Sigma_X^{-1}$ was divided by 3, to ensure the positive-definiteness of $\Sigma_X^{-1}$. We then set the diagonals of $\Sigma_X^{-1}$ as 1 and symmetrized it by averaging it with its transpose. The precision matrix $\Sigma_Y^{-1}$ was the same as $\Sigma_X^{-1}$ except that the connections of the top two hub nodes of $\Sigma_X^{-1}$ are multiplied by -1.
\item Simulation 3. In this simulation, we also considered data with block structures. Each block was a $100\times 100$ matrix and we generated each block of $\Sigma_X^{-1}$ and $\Delta$ randomly. Specifically, $60\%$ elements of each block of $\Sigma_X^{-1}$ were randomly chosen to be non-zero, and the non-zero elements were randomly sampled from $U(-0.1,0.1)$. We randomly selected 100 elements of the matrix $\Delta$ from $U(-0.5,0.5)$ (making sure $\Delta$ is symmetric). Then, $\Sigma_Y^{-1}$ was set as $\Sigma_X^{-1}+\Delta$. Lastly, we added a constant to their diagonal elements to make sure that $\Sigma_X^{-1}$ and $\Sigma_Y^{-1}$ are positive definite.
\end{itemize}

Figure \ref{Fig:simulation_ROC} illustrates the receiver operating characteristic curves (ROC) and the precision-recall curve of the three estimation methods for Simulation 1 with $p=100$ and $1000$ and $n=100$. Please see Supplementary for results of Simulation 2,3, and results of Simulation 1 with other choices of $p$ and $n$. Each point in the plots represent one value of the tuning parameter. If $\hat{\delta}_{jk}$ is the $(j,k)$th entry of a estimator $\hat{\Delta}$ and $\delta_{jk}^0$ is the $(j,k)$th entry of the true $\Delta$, the true positive (TP) and true negative (TN) rates are defined as
\begin{eqnarray*}
TP=\frac{\sum_{jk}I(\hat{\delta}_{jk}\neq 0 \ and \ \delta_{jk}^0\neq 0)}{\sum_{jk}I(\delta_{jk}^0\neq 0)},\quad TN=\frac{\sum_{jk}I(\hat{\delta}_{jk}=0 \ and\  \delta_{jk}^0=0)}{\sum_{jk}I(\delta_{jk}^0=0)},
\end{eqnarray*}
respectively. These simulations show that our D-trace loss estimator always had the highest AUC (Figure \ref{Fig:simulation_ROC} a,b and Supplementary Figure S1-S9 and Table S1-S6). Consistent with \citet{zhao2014direct}, the $l_1$-minimization method performed better than the fused graphical lasso in Simulation 2, but it performed similar to the fused graphical lasso in Simulation 1 and 3.  Precision-Recall curve also shows that, at the same level of TP rate, our D-trace estimator generally had a higher true discovery rate (TD) than the other two estimators (Figure \ref{Fig:simulation_ROC} c,d and Supplementary Figure S1-S9).

\begin{table}[!htpb]
\footnotesize
\centering
\caption{ Comparison of the 3 algorithms in terms of TP and TD rate in Simulation 1, n=100, 200, 500, respectively. The numbers in parentheses are the estimated standard deviations of TP and TD. The tuning parameters are either tuned with $L_F$-based BIC ($L_F$ columns) or $L_\infty$-based BIC ($L_\infty$ columns). DTL for the D-trace loss, FGL for the fused graphical lasso and L1-M for the $l_1$-minimization method in \citet{zhao2014direct}. \label{Tab:simulation_tablere}}
\begin{tabular}{cccccccccc}
&&\multicolumn{2}{c}{n=100}&&\multicolumn{2}{c}{n=200}&&\multicolumn{2}{c}{n=500}\\
&&$L_F$&$L_\infty$&
&$L_F$&$L_\infty$&
&$L_F$&$L_\infty$

\\
&&\multicolumn{7}{c}{p=100}\\
DTL&TP&5.7(3.0)&5.7(3.0)&&8.7(3.8)&8.7(3.8)&&15.4(4.8)&15.4(4.8)\\
&TD&78.0(15.9)&78.0(15.9)&&89.5(12.6)&89.5(12.6)&&97.6(3.9)&97.60(3.9)\\

FGL&TP&0.0(0.0)&0.0(0.0)&&0.0(0.0)&0.3(0.5)&&94.5(21.8)&0.7(1.0)\\
&TD&100.0(0.00)&99.0(10.0)&&100.00(0.0)&94.7(22.1)&&89.0(4.8)&97.3(14.7)\\

L1-M&TP&1.2(0.5)&1.2(0.6)&&1.3(0.4)&1.6(0.7)&&97.2(13.7)&2.00(1.5)\\
&TD&76.3(34.8)&75.5(34.5)&&88.3(23.0)&87.5(23.2)&&87.1(4.8)&97.9(11.2)\\
\\
&&\multicolumn{7}{c}{p=200}\\
DTL&TP&2.7(1.6)&2.7(1.6)&&4.3(2.1)&4.3(2.1)&&8.5(3.2)&8.5(3.2)\\
&TD&73.8(19.6)&73.8(19.6)&&89.9(8.0)&89.9(8.0)&&97.9(2.8)&97.9(2.8)\\

FGL&TP&0.0(0.0)&0.0(0.0)&&0.0(0.0)&0.2(0.3)&&0.0(0.0)&0.3(0.5)\\
&TD&100.0(0.0)&100.0(0.0)&&100.0(0.0)&92.0(27.3)&&100.0(0.0)&100.0(0.0)\\

&&\multicolumn{7}{c}{p=500}\\
DTL&TP&1.1(0.7)&1.1(0.7)&&2.0(1.0)&2.0(1.0)&&3.8(1.3)&3.8(1.3)\\
&TD&76.3(15.8)&76.3(15.8)&&87.3(9.4)&87.3(9.4)&&97.6(2.8)&97.6(2.8)\\

FGL&TP&0.0(0.0)&0.0(0.0)&&0.0(0.0)&0.1(0.1)&&0.0(0.0)&0.10(0.2)\\
&TD&100.0(0.0)&100.0(0.0)&&100.0(0.0)&89.0(31.5)&&100.0(0.0)&99.3(4.7)\\
\\

&&\multicolumn{7}{c}{p=1000}\\
DTL&TP&0.6(0.3)&0.6(0.3)&&1.0(0.5)&1.0(0.5)&&2.0(0.8)&2.0(0.8)\\
&TD&71.9(15.9)&71.9(15.9)&&88.3(9.8)&88.3(9.8)&&97.9(2.9)&97.9(2.9)\\

FGL&TP&0.0(0.0)&0.0(0.0)&&0.0(0.0)&0.0(0.0)&&0.00(0.00)&0.1(0.1)\\
&TD&100.0(0.0)&100.0(0.0)&&100.0(0.0)&90.0(30.2)&&100.0(0.0)&97.3(14.8)\\
\end{tabular}
\end{table}

We further studied the TP rates and TD rates of the three algorithms (Table \ref{Tab:simulation_tablere} and Supplementary Table S7,S8). We see that in most cases, our D-trace estimator had the highest TP rates with relatively high TD rates. With the parameters tuned by BIC, the TD rates remain to be high in most case, but TP rates are relatively low when $n$ is small (say $n=100$). This is probably due to the fact that the simulations are difficult for $n=100$. There are $p(p+1)/2$ parameters to estimate in the simulations. Even when $p=100$, the number of parameters are 5050, far larger than the number of observations. When we increased the number of observations, the TP rate could be significantly increased (Table \ref{Tab:simulation_tablere} and Supplementary Table). We also compared the computation time of each method (Supplementary Table S9-S11). As expected, we see that our algorithm took only a fraction of computation time of the $l_1$-minimization when $p$ is just 100. Our algorithm is computationally less efficient but comparable to the fused graphical lasso.

\section{Real Data}

In this section, we apply the D-trace loss estimation method and the other 2 methods to a gene expression data in colorectal cancer patients \citep{cancer2012comprehensive}. We are interested in study the gene regulatory network difference between the microsatellite instability (MSI) colorectal cancers and microsatellite stable (MSS) colorectal cancers. MSI cancer has a hypermutation phenotype resulted from impaired DNA mismatch repair (MMR). The MMR pathway includes genes such as \textit{MLH1}, \textit{MSH2} and \textit{MSH3} \citep{boland2010microsatellite}. The dyfunction of MMR pathway can be caused by mutations in the MMR genes or by the hypermethylation of MMR genes \citep{boland2010microsatellite}. In contrast to the MSI cancer genomes, MSS cancer genomes typically have more copy number variations but relatively less mutations. We therefore decide to see if there is any difference in gene regulatory network between the MSI and the MSS. The gene set we used is the colorectal cancer pathway as available in the KEGG pathway database (\citealp{749186}; \citealp{kanehisa2011kegg}). The genes in this pathway are known to play important roles in carcinogenesis of colorectal caners and there are 62 genes in this pathway. The gene expression data was downloaded from The Cancer Genome Atlas (TCGA) and we only used the patients with available MSI status information. This gave us 77 MSI patients and 122 MSS patients.

\begin{figure}[!htpb]
\centering
\subfigure[]{
\includegraphics[angle=0,width=4.5cm]{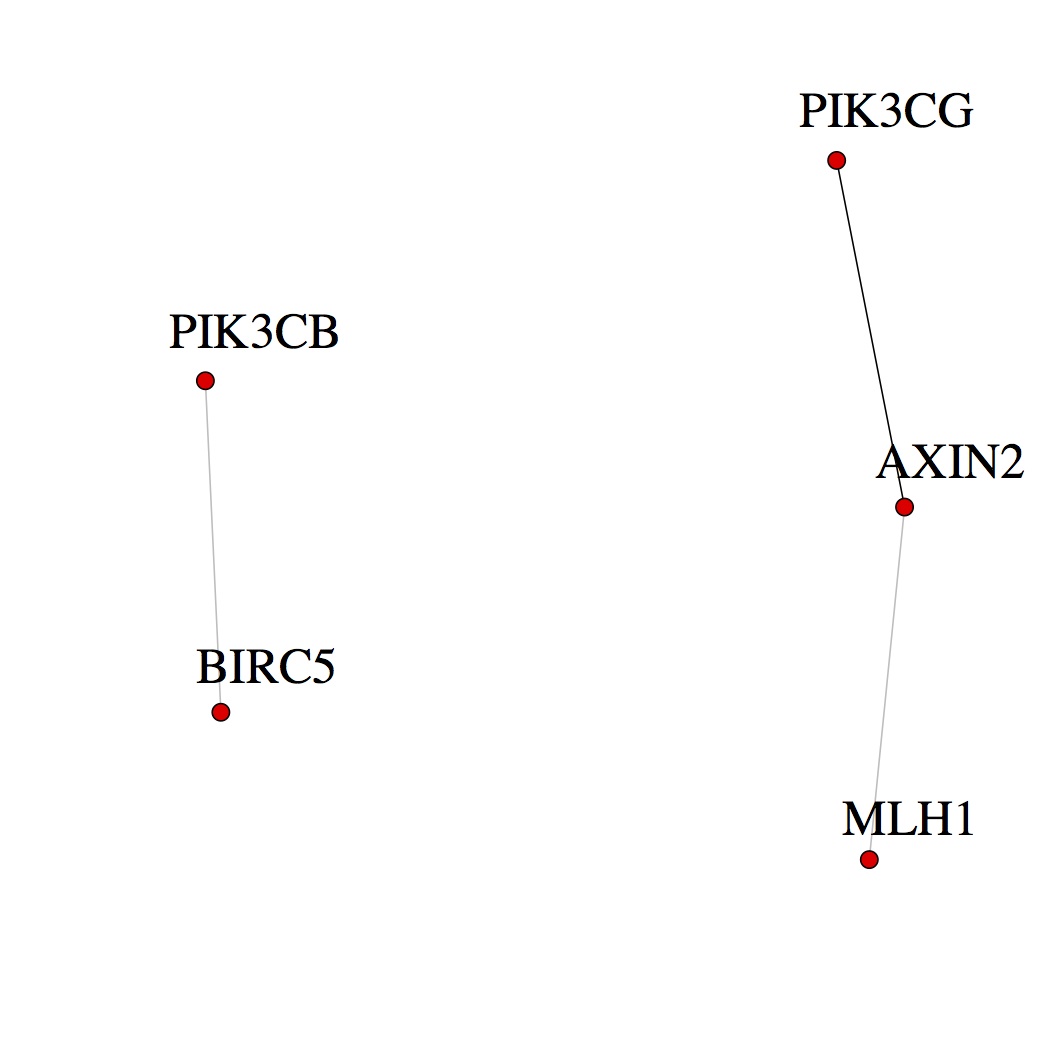}}
\subfigure[]{
\includegraphics[angle=0,width=4.5cm]{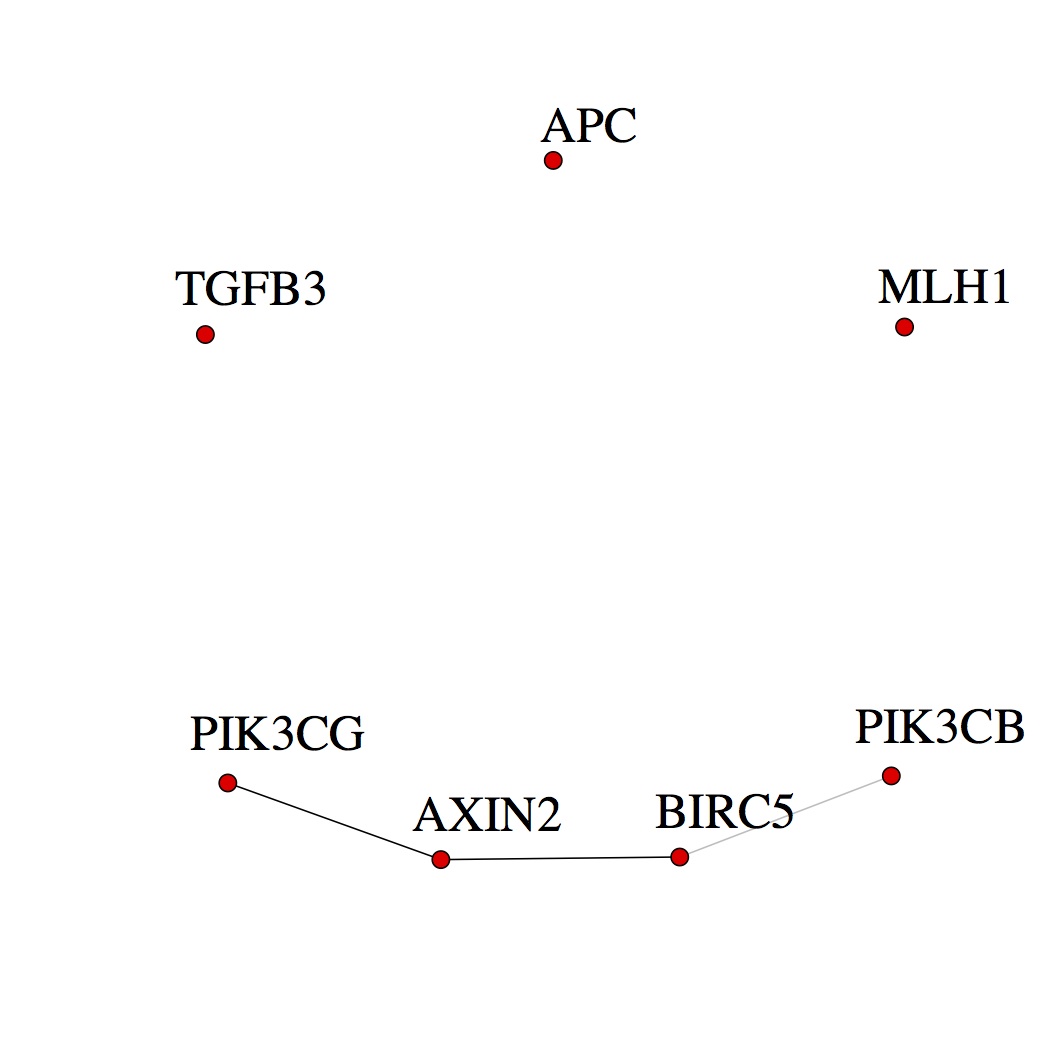}}
\subfigure[]{
\includegraphics[angle=0,width=4.5cm]{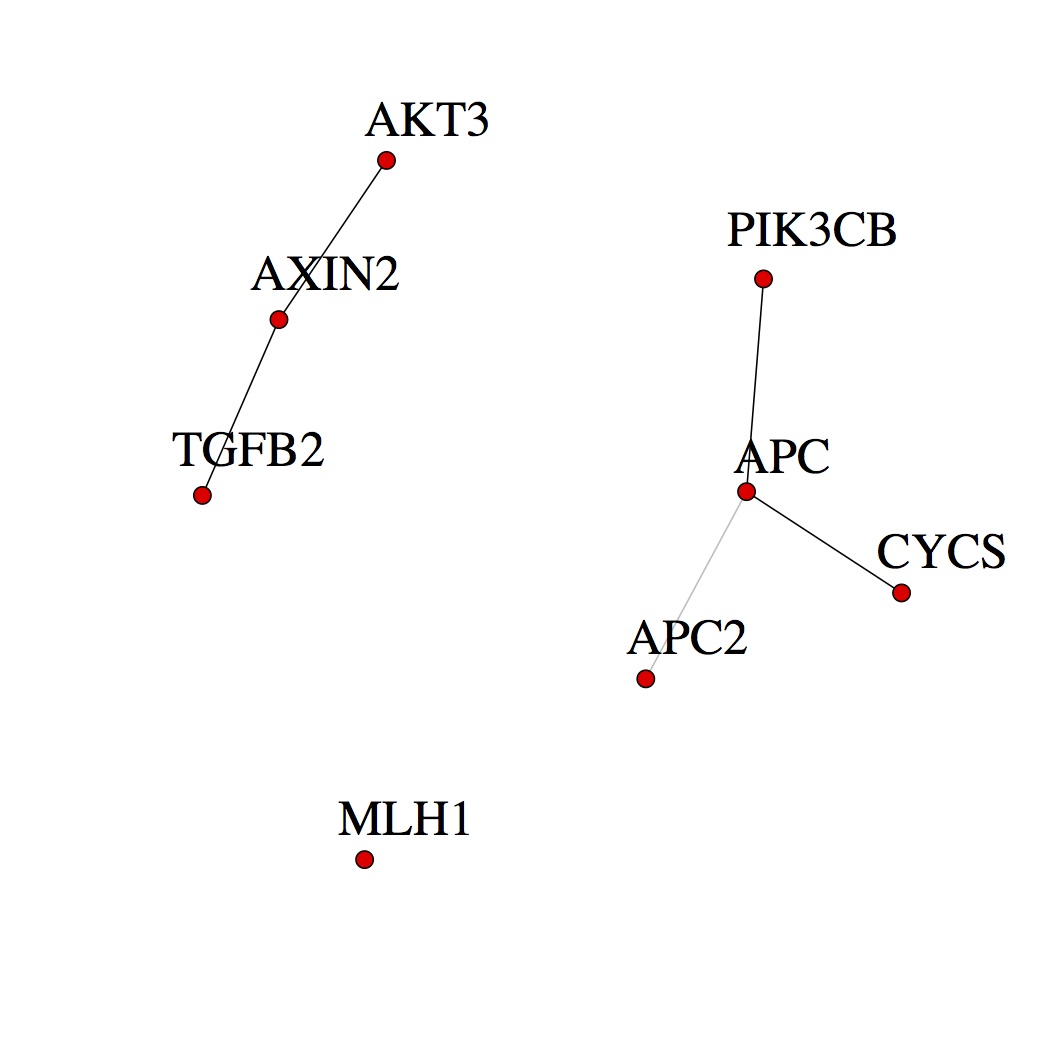}}
\caption{Gene regulatory network difference between the MSS and MSI colorectal cancers (Colorectal cancer pathway genes). (a) The D-trace loss, (b) the fused graphical lasso and (c) the $l_1$-minimization method. \label{figureMSSvsMSI}}
\end{figure}

Figure \ref{figureMSSvsMSI} shows the estimates given by the 3 methods with the tuning parameters tuned under $L_F$-norm. Here, we only show results of these methods under $L_F$-norm, since their estimates of $\Delta^*$ under $L_\infty$-norm and $L_F$-norm are the same. The three methods detected three common genes including \textit{MLH1}, \textit{AXIN2} and \textit{PIK3CB}. The gene \textit{MLH1} is a member of MMR gene family and its role in MSI colorectal cancer is well-established \citep{boland2010microsatellite}. The \textit{AXIN2} gene plays an important role in the regulation of the stability of beta-catenin in the Wnt signaling pathway. \textit{AXIN2} is frequently mutated in colorectal cancer genomes as well as many other types of cancers \citep{kandoth2013mutational}. It was shown that mutations in \textit{AXIN2} are associated with colorectal cancer with defective MMR \citep{liu2000mutations}. Our analysis of somatic mutation in 199 patients also showed that mutations in \textit{AXIN2} are more enriched in MSI patients (pvalue = 0.005944; Fisher's test; 7 MSI patients and 1 MSS patient harbored \textit{AXIN2} mutations). Consistent with the previous result, patients with \textit{AXIN2} mutations also tend to have more somatic mutations than patients without a \textit{AXIN2} mutation (Supplementary Figure 10 a; Pvalue=$1.6\times 10^{-11}$). These imply that the acquired somatic mutations on \textit{AXIN2} might cause the alteration of the interactions of \textit{AXIN2} with other genes.

The D-trace loss and the fused graphical lasso also identified two more common genes, \textit{PIK3CG} and \textit{BIRC5}. The gene \textit{PIK3CG} is significantly mutated in multiple cancers \citep{kandoth2013mutational}. The mutations on \textit{PIK3CG} are more enriched in MSI patients (pvalue= 0.005905; 10 MSI patients and 3 MSS patients harbored \textit{PIK3CG} mutations). Interestingly, based on data from 3134 cancer patients \citep{kandoth2013mutational}, we found that patients with mutations on \textit{PIK3CG} have significantly more somatic mutations than patients without a \textit{PIK3CG} mutation (Supplementary Figure S10 a; Pvalue=$8\times10^{-7}$; the Mann-Whitney U test). Although we did not find any report about the role of \textit{PIK3CG} in MMR, these data showed that \textit{PIK3CG} might play a role in MMR or it might be associated with hypermutation of cancer genomes. Interestingly, the mean expression of \textit{PIK3CG} was not significantly changed between MSS and MSI patients (pvalue = 0.07; t-test), but the correlation between \textit{PIK3CG} and \textit{AXIN2} significantly changed between two classes of patients (Supplementary Figure S10 b).

\section{Discussion}
\label{sec4}
This D-trace loss function can be generalized to compare multiple precision matrices. For example, if there are $K$ classes and the covariance matrix of the $k$th class is $\Sigma_K$, the multiple-call D-trace loss function may be defined as $\sum_{j<k}L_D(\Delta_{jk},\Sigma_j,\Sigma_k)$, where $\Delta_{jk}$ represents the precision matrix difference between class $j$ and $k$. If the precision matrix depends on continuous variables, the current technique cannot be used for detecting whether and how the precision matrix depends on the continuous variable. To handle such situations, we need make more assumptions on the precision matrices. The Markov random field model as used in \cite{hou2014guilt} could be a promising model for such situations, where one can explicitly model the dependency of the interactions on the covariates.
\section*{Acknowledgement}
This work is supported by the National Key Basic Research Project of China (Nos.2015CB910303, 2015CB856000,2016YFA0502303), and the NSFC (Nos. 11471022, 31171262, 31428012,31471246). We thank Dr. Sihai Dave Zhao for sharing his R code that implements his alternating direction method (\citealp{zhao2014direct}). We thank Dr. Teng Zhang for his discussions about his asymptotic property in his lasso-penalized D-trace loss.

\section*{Appendix: Technical proofs}

\subsection*{Tail Conditions}
According to \citet{5354541}, if a mean-zero random vector $X$ has a sub-Gaussian tail or a polynomial tail, then $X$ satisfies the tail condition $\mathcal{T}(f,\upsilon_*)$, i.e.,
there exist a constant $\upsilon_*>0$ and a function $f:\mathds{N}\times(0,\infty)\rightarrow (0,\infty)$ such that
\begin{equation*}
\label{fenge}
pr(\mid \hat{\Sigma}^n_{i,j}-\Sigma_{i,j}^*\mid \geq\delta)\leq 1/f(n,\delta),\quad 1\leq i,j\leq p,\quad 0<\delta<1/\upsilon_*,
\end{equation*}
where $\Sigma^*$ is the covariance matrix of $X$ and $\hat{\Sigma}^n$ is the sample covariance matrix given $n$ samples.
The function $f(n,\delta)$ is often monotonically increasing in $n$ and $\delta$ and continuous in $\delta$ (e.g. for distributions of sub-Gaussian tail or polynomial tail). Then, for each fixed $\delta>0$ and $n$, we can define the inverse functions for $r\geq 1$
\begin{equation*}
n_f(\delta,r)=argmax\{n: f(n,\delta)\leq r\}, \ \ \delta_f(n,r)=argmax\{\delta: f(n,\delta)\leq r\}.
\end{equation*}
\begin{remark} \label{note1}For any $0<\delta<1/\upsilon_*$ and $r\geq 1$, if $n>n_f(\delta,r)$, we have $f(n,\delta) > r$ and hence $\delta_f(n,r) < \delta$ since $f(n,\delta)$ is monotonically increasing in $\delta$. Thus,
$pr\{\mid \hat{\Sigma}^n_{i,j}-\Sigma_{i,j}^*\mid \geq\delta_f(n,r)\}\leq 1/f\{n,\delta_f(n,r)\} = r^{-1}$ because $f(n,\delta)$ is continuous in $\delta$ and $pr\{\|\hat{\Sigma}^n_X-\Sigma^*_X\|_\infty < \delta_f(n,r)\}> 1- p^2r^{-1}$.
\end{remark}
\noindent If $X$ is sub-Gaussian with parameter $\sigma$, we have (\citealp{5354541})
\begin{eqnarray}\label{subGaussianTail}
\upsilon_{*}&=&\{\max_i\Sigma^*_{i,i}8(1+4\sigma^2) \}^{-1},\nonumber\\
f(n,\delta)&=&\exp(c_{*}n\delta^2)/4~~\mbox{with}~~c_{*}=\{128(1+4\sigma^2)^2\max_i(\Sigma^*_{i,i})^2\}^{-1},\nonumber\\
\delta_{f}(n,p^\eta)&=&\{128(1+4\sigma^2)^2\max_i(\Sigma^*_{i,i})^2(\eta \log p+\log 4)/n\}^{1/2},\\
n_{f}(\delta,p^{\eta})&=&128(1+4\sigma^2)^2\max_i(\Sigma^*_{i,i})^2(\eta \log p+\log 4)/\delta^2.\nonumber
\end{eqnarray}
If $X$ has a polynomial tail with parameters $m$ and $K_m$, we have (\citealp{5354541}, Section 2.3.2)
 \begin{equation}\label{polynomialTail}
 \begin{aligned}
&\upsilon_{*}=0,~~
f(n,\delta)=c_{*}n^m\delta^{2m}~~\mbox{with}~~c_{*}=2^{-2m}(\max_i\Sigma^*_{i,i})^{-2m}(K_m+1)^{-1},\nonumber\\
&\delta_{f}(n,p^{\eta})=p^{\eta/(2m)}c_{*}^{-1/(2m)}n^{-1/2},~~
n_{f}(\delta,p^{\eta})=p^{\eta/m}c_{*}^{-1/m}\delta^{-2}.
\end{aligned}
\end{equation}

For any subset T of $\{1,\ldots,p\}\times\{1,\ldots,p\}$, we denote by $\mbox{vec}(\Gamma)_T$ the sub-vector of $\mbox{vec}(\Gamma)$ made up of elements of $\Gamma_T$. We further define
\begin{equation*}
\begin{aligned}
&\epsilon=\|\hat{\Sigma}_X-\Sigma_X^*\|_\infty\|\hat{\Sigma}_Y-\Sigma_Y^*\|_\infty+\|\Sigma^*_X\|_\infty\|\hat{\Sigma}_Y-\Sigma^*_Y\|_\infty+\|\Sigma^*_Y\|_\infty\|\hat{\Sigma}_X-\Sigma^*_X\|_\infty,\\
&\Delta_\Gamma=\hat{\Gamma}-\Gamma^*,\quad \Delta_\Sigma=\hat{\Sigma}_X-\hat{\Sigma}_Y-\Sigma_X^*+\Sigma_Y^*,\quad \tilde{\epsilon}=\|\Delta_\Sigma\|_\infty.
\end{aligned}
\end{equation*}

We first present two lemmas for proving the main theorems. The proofs of the lemmas are given in Supplementary.
\begin{lemma}
\label{lemma1}
 Define $\hat{\Delta}$ by
\begin{equation}
\label{check}
\hat{\Delta}=argmin_{\Delta} L_D(\Delta, \hat{\Sigma}_X, \hat{\Sigma}_Y)+\lambda_n\|\Delta\|_{1}
\end{equation}
\begin{itemize}
\item[(a)]
Then $vec(\hat{\Delta})_{S^c}$=0 if
 \begin{equation}
 \label{cona}
 \begin{aligned}
\max_{e\in S^c}\|\hat{\Gamma}_{e,S}(\hat{\Gamma}_{S,S})^{-1}\|_1\leq 1-\alpha/2,~  \|\hat{\Gamma}_{e,S}\hat{\Gamma}_{S,S}^{-1}-\Gamma^*_{e,S}\Gamma_{S,S}^{*-1}\|_1<\frac{\alpha\lambda_n}{8M},~ \tilde{\epsilon}\leq \frac{\alpha\lambda_n}{2(4-\alpha)}.
  \end{aligned}
 \end{equation}
 \item[(b)] $vec(\hat{\Delta})_{S^c}$=0 if
\begin{equation}
\label{eplison}
\epsilon<\frac{1}{6s\kappa_{\Gamma}},
\end{equation}
\begin{equation}
\label{conb2}
3s\epsilon(\kappa_{\Gamma}+2sM^2\kappa_{\Gamma}^2)\leq 0.5\alpha \min(1,0.25\lambda_nM^{-1}),
\end{equation}
\begin{equation}
\label{conb3}
 \tilde{\epsilon}\leq \frac{\alpha\lambda_n}{2(4-\alpha)}.
\end{equation}
\item[(c)]
 Assuming the conditions in part (b), then we also have
\begin{equation}
\label{conc}
\|\hat{\Delta}-\Delta^*\|_\infty<(\tilde{\epsilon}+\lambda_n)\kappa_\Gamma+3(\tilde{\epsilon}+2M+\lambda_n)s\epsilon\kappa_\Gamma^2
\end{equation}
\end{itemize}
\end{lemma}

\begin{lemma}
\label{lemma2}
Assuming (\ref{eplison}), we have
\begin{equation}
\label{R}
\|R(\Delta_{\Gamma})\|_{1,\infty}\leq 6s^2\epsilon^2\kappa_{\Gamma}^3,\ \ \ \|R(\Delta_{\Gamma})\|_\infty\leq 6s\epsilon^2\kappa_{\Gamma}^3,
\end{equation}
where $R(\Delta_{\Gamma})=\{\Gamma_{S,S}^*+(\Delta_{\Gamma})_{S,S}\}^{-1}-\Gamma_{S,S}^{*-1}+\Gamma_{S,S}^{*-1}(\Delta_{\Gamma})_{S,S}\Gamma_{S,S}^{*-1}$. Moreover, we also have
\begin{equation}
\label{l1wuq}
\|\hat{\Gamma}_{S,S}^{-1}-\Gamma_{S,S}^{*-1}\|_{1,\infty}\leq 6s^2\epsilon^2\kappa_{\Gamma}^3+2s\epsilon\kappa_{\Gamma}^2,
\end{equation}
\begin{equation}
\label{lwuq}
\|\hat{\Gamma}_{S,S}^{-1}-\Gamma_{S,S}^{*-1}\|_\infty\leq 6s\epsilon^2\kappa_{\Gamma}^3+2\epsilon\kappa_{\Gamma}^2.
\end{equation}
\end{lemma}

\subsection*{Proof of Theorem 1 and Theorem 3}

\begin{proof} Since $X$ has a sub-Gaussian tail or a polynomial tail, we have $X$ satisfies the tail condition $\mathcal{T}(f_X,\upsilon_{X*})$, where $f_X$ and $\upsilon_{X*}$ as defined in (\ref{subGaussianTail}) or (\ref{polynomialTail}) with $\Sigma^*$ replaced by $\Sigma^*_X$. Similarly, $Y$ also satisfies the tail condition $\mathcal{T}(f_Y,\upsilon_{Y*})$. If we take $\upsilon_{*} = \max(\upsilon_{X*},\upsilon_{Y*})$, then $X$ and $Y$ also satisfy the tail condition $\mathcal{T}(f_X,\upsilon_{*})$ and $\mathcal{T}(f_Y,\upsilon_{*})$, respectively. Let
$$\bar{\delta} = \min\bigg\{-M+\sqrt{M^2+(6s\kappa_\Gamma)^{-1}},-M+\sqrt{2M^2+\frac{\alpha}{24s(2sM^2\kappa_\Gamma^2+\kappa_\Gamma)}}, \frac{\alpha M}{4-\alpha},1/\upsilon_*\bigg\}.$$
For sub-Gaussian-tailed distribution, we have
\begin{equation*}
\begin{aligned}
 \bar{\delta} = &\min\bigg\{-M+\sqrt{M^2+(6s\kappa_\Gamma)^{-1}},-M+\sqrt{M^2+{\alpha}/\left\{24s(2sM^2\kappa_\Gamma^2+\kappa_\Gamma)\right\}}, \\
 &(\alpha M)/(4-\alpha),\min_{X,Y}\{\max_i\Sigma^*_{X,i,i}8(1+4\sigma_X^2),\max_i\Sigma^*_{Y,i,i}8(1+4\sigma_Y^2)\}\bigg\}.
\end{aligned}
\end{equation*}
 For polynomial-tailed distribution, we have $$\bar{\delta} = \min\bigg\{-M+\sqrt{M^2+(6s\kappa_\Gamma)^{-1}},-M+\sqrt{M^2+{\alpha}/\left\{24s(2sM^2\kappa_\Gamma^2+\kappa_\Gamma)\right\}}, \frac{\alpha M}{4-\alpha}\bigg\}.$$
In the following, for $\eta>2$, we assume $n_X>n_{f_X}(\bar{\delta},p^\eta)$, $n_Y>n_{f_Y}(\bar{\delta},p^\eta)$
and $$\lambda_n=\max\bigg\{2(4-\alpha)(\delta_{f_X}+\delta_{f_Y})/\alpha,{24sM(2sM^2\kappa_\Gamma^2+\kappa_\Gamma)(\delta_{f_X}\delta_{f_Y}+M\delta_{f_Y}+M\delta_{f_X})}/{\alpha}\bigg\},$$
where $\delta_{f_X}=\delta_{f_X}(n_X,p^\eta)$ and $\delta_{f_Y}=\delta_{f_Y}(n_Y,p^\eta)$.

(a) We first prove the first inequalities of Theorem 1 and 3 using Lemma \ref{lemma1}. From Fact \ref{note1}, for $n_X>n_{f_X}(\bar{\delta},p^\eta)$, with probability at least $1-1/p^{\eta-2}$, we have
\begin{equation*}
\begin{aligned}
\label{trueep}
\|\hat{\Sigma}^{n_X}_X-\Sigma^*_X\|_\infty\leq&\delta_{f_X}(n_X,p^\eta)<\bar{\delta}.
\end{aligned}
\end{equation*}
Similar result also holds for $Y$ with $n_Y>n_{f_Y}(\bar{\delta},p^\eta)$. Now we show that the 3 conditions in Lemma \ref{lemma1} (b) are satisfied. Since $\|\hat{\Sigma}^{n_X}_X-\Sigma^*_X\|_\infty<\bar{\delta}$, $\|\hat{\Sigma}^{n_Y}_Y-\Sigma^*_Y\|_\infty<\bar{\delta}$ and $\bar{\delta}\leq \big(-2M+\sqrt{4M^2+2/(3s\kappa_\Gamma)}~\big)/2$, the condition (\ref{eplison}) can be easily verified by using some algebra.

Since $\delta_{f_X}=\delta_{f_X}(n_X,p^\eta) < \bar{\delta} \leq {\alpha M}/{(4-\alpha)}$ (similar for $Y$), we have
${2(4-\alpha)}(\delta_{f_X}+\delta_{f_Y})/{\alpha}\leq 4M.$
From
$\delta_{f_X} < \bar{\delta} \leq -M+\sqrt{M^2+\alpha/\left\{24s(2sM^2\kappa_\Gamma^2+\kappa_\Gamma)\right\}},$
we get
$\delta_{f_X}\delta_{f_Y}+M\delta_{f_Y}+M\delta_{f_X}\leq {\alpha}/\left\{24s(2sM^2\kappa_\Gamma^2+\kappa_\Gamma)\right\}.$
Then, by the definition of $\lambda_n$, we have $0.25M^{-1}\lambda_n\leq 1$. Furthermore, we have
$\epsilon\leq\delta_{f_X}\delta_{f_Y}+M\delta_{f_Y}+M\delta_{f_X}\leq{\alpha}/\left\{24s(2sM^2\kappa_\Gamma^2+\kappa_\Gamma)\right\},$
which implies that
$3s\epsilon(\kappa_{\Gamma}+2sM^2\kappa_{\Gamma}^2)\leq \alpha/6 < 0.5\alpha$
and
\begin{eqnarray*}
3s\epsilon(\kappa_{\Gamma}+2sM^2\kappa_{\Gamma}^2)&\leq& 3s (\kappa_{\Gamma}+2sM^2\kappa_{\Gamma}^2) (\delta_{f_X}\delta_{f_Y}+M\delta_{f_Y}+M\delta_{f_X})\\
&=& 8^{-1}\alpha M^{-1} \frac{24sM(\kappa_{\Gamma}+2sM^2\kappa_{\Gamma}^2) (\delta_{f_X}\delta_{f_Y}+M\delta_{f_Y}+M\delta_{f_X})}{\alpha}\\
&\leq& 8^{-1}\alpha M^{-1}\lambda_n.
\end{eqnarray*}
Combining the above two results, we can obtain (\ref{conb2}). For the condition (\ref{conb3}), we have
\begin{eqnarray*}
\tilde{\epsilon}&\leq& \|\hat{\Sigma}^{n_X}_X-\Sigma^*_X\|_\infty + \|\hat{\Sigma}^{n_Y}_Y-\Sigma^*_Y\|_\infty \leq \delta_{f_X}+\delta_{f_Y} \leq \frac{\alpha\lambda_n}{2(4-\alpha)}.
\end{eqnarray*}
Then, by Lemma \ref{lemma1} (c),
\begin{equation}
\label{endinfty}
\begin{aligned}
 \|\hat{\Delta}-\Delta^*\|_\infty<&(\tilde{\epsilon}+\lambda_n)\kappa_\Gamma+3(\tilde{\epsilon}+2M+\lambda_n)s\epsilon\kappa_\Gamma^2 \\
\leq &(\delta_{f_X}+\delta_{f_Y}+\lambda_n)\kappa_\Gamma\\
&+3s\kappa_\Gamma^2(\delta_{f_X}+\delta_{f_Y}+\lambda_n+2M)(\delta_{f_X}\delta_{f_Y}+M\delta_{f_Y}+M\delta_{f_X})\\
\leq &\{\kappa_\Gamma+3s\kappa_\Gamma^2(\delta_{f_X}\delta_{f_Y}+M\delta_{f_Y}+M\delta_{f_X})\} (\delta_{f_X}+\delta_{f_Y}+\lambda_n)\\
&+6sM\kappa_\Gamma^2(\delta_{f_X}\delta_{f_Y}+M\delta_{f_Y}+M\delta_{f_X}).
 \end{aligned}
\end{equation}
Let $A={\alpha M}/{(4-\alpha)}$, then $\delta_{f_X}<A$ and $\delta_{f_Y}<A$.

Suppose that $X$ and $Y$ are sub-Gaussian. We have (similar inequality also holds for $\delta_{f_Y}$)
\begin{eqnarray*}
\delta_{f_X}&\leq&\{128(1+4\sigma_X^2)^2M^2\}^{1/2}\{(\eta \log p+\log 4)/n_X\}^{1/2}\\
&\leq&C_{G}^{1/2}\{(\eta \log p+\log 4)/\min(n_X,n_Y)\}^{1/2},
\end{eqnarray*}
where $C_{G}=128\{1+4\max(\sigma_X^2,\sigma_Y^2)\}^2M^2$. By the definition of $\lambda_n$ and (\ref{endinfty})
\begin{equation*}
\begin{aligned}
 & \|\hat{\Delta}-\Delta^*\|_\infty \leq (C_{G1}+C_{G2})\bigg\{\frac{\eta \log p+\log 4}{\min(n_X, n_Y)}\bigg\}^{1/2},\\
 \end{aligned}
\end{equation*}
where
\begin{equation}\label{C1C2Definition}
\begin{aligned}
C_{G1}=&\{ \kappa_\Gamma+3s\kappa_\Gamma^2(A^2+2MA)\}\bigg[2C_{G}^{1/2}\\
&+\max\{24sM(2sM^2\kappa_\Gamma^2+\kappa_\Gamma)(2MC_{G}^{1/2}+AC_{G}^{1/2})/\alpha,{4C_{G}^{1/2}(4-\alpha)}/{\alpha}\}\bigg],\\
C_{G2}=&6sM\kappa_\Gamma^2(2MC_{G}^{1/2}+AC_{G}^{1/2}).
\end{aligned}
\end{equation}

Suppose that $X$ and $Y$ are of polynomial tails. Let $C_{P}=4M^{2}\{\max(K_{Xm},K_{Ym})+1\}^{1/m}$. Thus,
$\delta_{f_X}\leq C_{P}^{1/2}p^{\eta/(2m)}\min{(n_X, n_Y)}^{-1/2}$
and
\begin{equation*}
\begin{aligned}
 & \|\hat{\Delta}-\Delta^*\|_\infty  \leq (C_{P1}+C_{P2})p^{\eta/(2m)}\min{(n_X, n_Y)}^{-1/2},\\
 \end{aligned}
\end{equation*}
with $C_{P1}$ and $C_{P2}$ as defined in (\ref{C1C2Definition}) with $C_{G}$ replaced by $C_{P}$.

Define $M_G=C_{G1}+C_{G2}$ and $M_P=C_{P1}+C_{P2}$ and we have proved the first inequalities in Theorem 1 and 3.

(b) We now prove the second inequalities of Theorem 1 and 3. The above proof showed that the 3 conditions in Lemma \ref{lemma1} (b) are satisfied and thus the nonzero elements of $\hat{\Delta}$ is a subset of the nonzero elements of $\Delta^*$. Thus,
\begin{equation*}
\label{endF}
\begin{aligned}
\|\hat{\Delta}-\Delta^*\|_F\leq &s^{1/2}\|\hat{\Delta}-\Delta^*\|_\infty \leq &(C_{G1}+C_{G2})\bigg\{\frac{\eta \log p+\log 4}{\min{(n_X, n_Y)}}\bigg\}^{1/2}s^{1/2}.
\end{aligned}
\end{equation*}

\end{proof}

\subsection*{Proof of Theorem 2 and Theorem 4}
\begin{proof}
We only prove the sub-Gaussian case since the proof of the polynomial case is similar. From Theorem 1, we have
\begin{equation*}
\mid \tilde{\Delta}_{i,j}-\Delta^*_{i,j}\mid \leq (C_{G1}+C_{G2})\bigg\{\frac{\eta \log p+\log 4}{\min{(n_X, n_Y)}}\bigg\}^{1/2}.
\end{equation*}
By the proof of Theorem 1, we know that the nonzero elements of $\hat{\Delta}$ is a subset of the nonzero elements of $\Delta^*$. Given the conditions in Theorem 2, these implies that sgn($\tilde{\Delta}_{i,j}$)=sgn($\Delta^*_{i,j}$) for all $i,j$ with probability at least $1-2/p^{\eta-2}$. The conclusion of Theorem 2 is thus followed.
\end{proof}

\bibliographystyle{biometrika}
\bibliography{Refrevised}

\end{document}